\documentclass[showpacs,preprintnumbers,nofootinbib]{revtex4}
\usepackage{amsmath,amsfonts,amssymb,amsbsy,bbm}
\begin{document}

\vspace*{1cm}

\title{The Isometries of Low-Energy Heterotic M-Theory\\}

\author{Edmund J. Copeland}
\email{ed.copeland@nottingham.ac.uk}
\affiliation{School of Physics and Astronomy,
University of Nottingham, University Park, Nottingham, NG7 2RD, UK}
\author{James Ellison}
\email{j.b.ellison@sussex.ac.uk}
\affiliation{Department of Physics and Astronomy,
University of Sussex, Brighton, BN1 9QJ, UK}
\author{Andr\'e Lukas}
\email{lukas@physics.ox.ac.uk}
\affiliation{Rudolf Peierls Centre for Theoretical Physics,
University of Oxford, Oxford OX1 3NP, UK}
\author{Jonathan Roberts}
\email{j.roberts@sussex.ac.uk}
\affiliation{Department of Physics and Astronomy, University
of Sussex, Brighton, BN1 9QJ, UK}

\date{\today}

\begin{abstract}
\noindent We study the effective $D=4$, $\mathcal{N}=1$ supergravity description of
five-dimensional heterotic \mbox{M-theory} in the presence of an M5 brane, and derive the
Killing vectors and isometry group for the K\"ahler moduli-space metric. The group is found
to be a non-semisimple maximal parabolic subgroup of $Sp(4,\mathbb{R})$, containing a
non-trivial $SL(2,\mathbb{R})$ factor. The underlying moduli-space is then naturally
realised as the group space $Sp(4,\mathbb{R})/U(2)$, but equipped with a nonhomogeneous
metric that is invariant only under that maximal parabolic group. This nonhomogeneous metric
space can also be derived via field truncations and identifications performed on
$Sp(8,\mathbb{R})/U(4)$ with its standard homogeneous metric. In a companion paper we use
these symmetries to derive new cosmological solutions from known ones.

\end{abstract}


%specifies that the paper contains material related to M-theory, compactifications,
%Lie algebras, Riemannian geometries and complex manifolds

\maketitle

%===============================================INTRODUCTION==========================================
%=====================================================================================================

\section{Introduction}

It has been known for some time that the classical supergravity actions often possess scalar
fields with highly constrained interactions. In particular, the $D=4, \mathcal{N}=1$
theories have scalar field kinetic terms that must collectively arise from the K\"ahler
potential of a complex K\"ahler manifold~\cite{Zumino:1979et}. This means that the scalars
can be reinterpreted as real-valued coordinates on a complex space, and so their equations
of motion possess an underlying geometric significance. Therefore, when investigating the
scalar-field cosmology of supergravity models, it is often useful to identify the structure
of these scalar-field manifolds and their symmetries~\cite{Lidsey:1999mc}. Indeed, this
knowledge sometimes obviates the need to solve the full equations of motion of the system,
since repeated symmetry transformations on special solutions often ``builds'' the general
solution for us.

In this paper we will be considering a scalar-field cosmology that descends from
Horava-Witten (HW) theory coupled to an M5 brane~\cite{Horava:1996ma,Bandos:1997ui}. In
particular, we will consider a compactification of HW theory on a Calabi-Yau three-fold
$CY_3$, leading to the known form of five-dimensional heterotic
M-theory~\cite{Lukas:1998tt,Brandle:2001ts}. Within the context of this theory, it was shown
in Ref.~\cite{Copeland:2001zp} how a further compactification to four-dimensions leads to an
$\mathcal{N}=1$ supergravity theory with an unusual braneworld cosmology. For example, it
transpires that a new scalar corresponding to the M5 brane position must be included in the
set of cosmologically significant fields, and this leads to a forcing effect whereby the
ambient dimensions change size as the brane moves. Moreover, the frictional forces acting
back on the brane are such that it accelerates and then decelerates back to rest,  mimicking
a time-dependent force of finite duration. The internal dimensions of the model are thus
subjected to a brief period of brane-induced changes, but eventually act as free scalars
that are contracting or expanding with fixed rates.

However, not all of the scalar fields were considered in Ref.~\cite{Copeland:2001zp}, since
the axionic fields were consistently truncated away. This means that only a portion of the
full scalar manifold was explored, and that the M5 brane behaviour is liable to be only an
approximation once the axions are restored. Therefore, following on from the work of
Ref.~\cite{Copeland:2001zp}, we wish to analyse the structure and symmetries of the
\emph{full} K\"ahler metric, and also discuss the form of the underlying manifold on which
the metric is placed. In a companion paper~\cite{prep} we will then use this information to
find new solutions to the equations of motion.

We now summarise our results. We find that the isometry group of the K\"ahler metric is a
maximal parabolic subgroup of $Sp(4,\mathbb{R})$, which possesses an $SL(2,\mathbb{R})$
group of $T$-duality transformations. The underlying K\"ahler manifold is the Siegel plane
$SH_{2} \cong Sp(4,\mathbb{R})/U(2)$ equipped with a nonhomogeneous metric. We show this by
assembling the metric from the right-invariant one forms of the group
$Sp(4,\mathbb{R})/U(2)$. We also demonstrate how to derive the nonhomogeneous metric from a
homogeneous metric on the higher-dimensional Siegel plane $SH_{4}\cong
Sp(8,\mathbb{R})/U(4)$, by making certain field truncations and identifications.

%============================================SECTION 1===============================================================

\section{The Four-dimensional action}

We now review the $D=4, \mathcal{N}=1$ supergravity action presented in
Ref.~\cite{Copeland:2001zp}. Recall that this was derived via a compactification of 11D
supergravity on $S^1/\mathbb{Z}_{2} \times CY_{3}$, leading to two four-dimensional boundary
planes separated along a fifth dimension. A single M5 brane was also included in the space,
by wrapping it on a holomorphic 2-cycle of the $CY_{3}$. The brane then appears as a
three-brane of charge $q$ that lies parallel to the boundaries, and which can move along the
interval. Importantly, the interaction between the boundaries and brane leads to the
existence of a static, triple-domain wall BPS solution. One can then consider further
reducing on this solution, so as to find a supergravity theory describing slowly varying
fluctuations about the static BPS vacuum. This contains the six scalar fields
$\beta,\chi,\phi,\sigma,z,\nu$ with the following non-standard kinetic terms
\begin{align}
\begin{split}
S_{4}=-\frac{1}{2\kappa_{4}^2}\int_{M_{4}} d^{4}x\sqrt{-g}\left[
\frac{1}{2}R+\frac{3}{4}(\partial\beta)^{2}+3e^{-2\beta}(\partial\chi)^{2}+\frac{1}{4}(\partial\phi)^{2}+
\frac{1}{4}e^{-2\phi}\left(\partial\sigma + 4qz\partial\nu\right)^{2}
 \right.\\
\left.+\frac{1}{2}qe^{\beta-\phi}(\partial z)^{2}+2qe^{-\beta-\phi}(\partial\nu-\chi
\partial z)^{2} \right]\label{4daction}
\end{split}
\end{align}
Each of these scalars has an underlying significance in terms of the $D=5$ parent theory
from which it descends. The scalar $\beta$ is the zero-mode of the $g_{55}$ component in the
$D=5$ metric, and measures the separation between the boundaries. Specifically, the
separation is given by $\pi\rho e^{\beta}$ in terms of some dimensionful reference size
$\pi\rho$. The field $\phi$ represents the orbifold-averaged Calabi-Yau volume, such that
the physical size is given by $v e^{\phi}$ in terms of a dimensionful reference volume $v$.
The scalars $\sigma,\chi$ originate from the bulk three-form and graviphoton field
respectively. The field $z$ measures the position of the bulk brane between the boundaries,
and has been normalised such that $z \in (0,1)$. The points $z=0,1$ then correspond to the
boundaries themselves. Lastly, the field $\nu$ arises from the self-dual two-form on the
brane worldvolume.

This reduction on a BPS solution guarantees that the scalars must group into supersymmetric
multiplets described by a supersymmetric action. One can verify that they naturally fall
into the pairs $(\phi,\sigma)$,$(\beta,\chi)$,$(z,\nu)$, which are the bosonic components of
chiral superfields $S,T,Z$ as follows
\begin{align}\label{complexstructure}
S= e^{\phi} +qz^{2}e^{\beta} + i\left(\sigma +2qz^2\chi\right) \quad ,\quad T= e^{\beta}
+2i\chi \quad,\quad Z= e^{\beta}z + 2i(-\nu+ z\chi)
\end{align}
This naturally leads to a K\"ahler manifold expression for the scalar part of the action
\begin{align}
S_{4}=-\frac{1}{2\kappa_{4}^{2}}\int d^{4}x \sqrt{-g} \left(\frac{1}{2}R+
K_{ij}\:\partial_{\mu}{\Phi}^{i} \partial^{\mu}{\bar{\Phi}}^{\bar{j}}\right)
\end{align}
where the superfields are grouped into a coordinate vector $\Phi=(S,T,Z)$, with the complex
conjugate coordinates denoted by $\bar{\Phi}$. The K\"ahler metric $K_{i\bar{j}}$ is given
by
\begin{align}
K_{i\bar{j}} = \frac{ \partial^{2}{K} }{\partial{\Phi^{i}}\partial{\bar{\Phi}^{\bar{j}}}}
\end{align}
in terms of the K\"ahler potential
\begin{align}\label{kahlerpot}
K = -\ln\left[S+\overline{S}-q\frac{(Z+\overline{Z})^{2}}{T+\overline{T}}\right]
-3\ln\left(T+\overline{T}\right)
\end{align}

%==================================SECTION 2========================================================================

\section{Killing vectors and Isometry group}

To understand the scalar-field dynamics of the action Eq.~\eqref{4daction}, it is important
to study the isometries of the associated K\"ahler metric. The rationale for this is as
follows. Usually the underlying scalar-field manifold is a group $G/H$ where $G$ is a Lie
group of diffeomorphisms that acts transitively on $G/H$, and $H$ is a normal subgroup of
$G$~\footnote{If $H$ is not a normal subgroup, then the coset $G/H$ is a manifold with no
group structure. Thus, not every coset $G/H$ is a group, and care should be taken to
distinguish between ``cosets'' and ``cosets that are also groups''. In the following we will
always consider spaces of the form $G/H$ where $H$ is normal in $G$, and so these spaces are
automatically Lie groups in their own right.}. Moreover, the K\"ahler metric on $G/H$ is
often invariant under some isometry group $G\:' \subseteq G$. The natural consequence of
this is as follows. If we are given any solution to the scalar-field equations of motion,
then this can be transformed into a new and generally more complicated solution by
systematically applying an isometry transformation in $G\:'$. Although only $G$ is strictly
transitive on the underlying group space $G/H$, so that the \emph{general} solution cannot
be built unless $G'=G$, we can nonetheless make significant progress by applying isometries.
In particular, we can find complicated new solutions that might otherwise be inaccessible
using conventional solving techniques.

Based on this, we consider the isometries of Eq.~\eqref{4daction}, which have not yet been
discussed in the braneworld literature. To this end, we recognise that the action
Eq.~\eqref{4daction} is written in the Einstein frame, such that the gravitational
contribution is cleanly decoupled from the scalar fields. Consequently, the spacetime
manifold will transform as a singlet under the isometry group of the scalar manifold, and so
can be considered some fixed background that does not affect the permissible scalar field
symmetries. Therefore, from now on we ignore the Ricci scalar term in the action. For
simplicity, we will present our results in term of the scalar fields
$\beta,\chi,\phi,\sigma,z,\nu$, so that the scalar-field manifold is treated as real and
six-dimensional. Where necessary, we will comment upon the analogous complex results in
terms of the natural complex fields $S,T,Z$.

The infinitesimal isometries of the metric Eq.~\eqref{4daction} are generated by
left-invariant Killing vector fields $L^{i}$. By solving the Killing equations,
\begin{align*}
\nabla_{i}L_{k} + \nabla_{k}L_{i} = 0
\end{align*}
one can verify that there are seven real Killing vector fields as follows
\begin{align*}
L^{1} &= \partial_{\beta} +\chi\partial_{\chi} -\frac{1}{2}\:z\partial_{z}
 +\frac{1}{2}\:\nu\partial_{\nu} \\
L^{2} &= -2\chi\partial_{\beta}+\left(\frac{1}{4}e^{2\beta}-\chi^{2}\right)\partial_{\chi} +
\nu\partial_{z}-2q\nu^2\partial_{\sigma}
 \\
L^{3} &= \partial_{\chi} + z\partial_{\nu} -2qz^{2}\partial_{\sigma} \\
L^{4} &= \partial_{\phi} + \frac{1}{2}\:z\partial_{z} +
\sigma\partial_{\sigma} +\frac{1}{2}\:\nu\partial_{\nu} \\
L^{5} &= \partial_{z} -4q\nu \partial_{\sigma} \\
L^{6} &= 4q \partial_{\sigma}\\
L^{7} &= \partial_{\nu}
\end{align*}
We note that each of these real vector fields also corresponds to a \emph{holomorphic}
Killing vector field of the complex manifold with coordinates $S,T,Z$. That is, every such
field can be decomposed as the cleanly-separated sum
\begin{align*}
L^{a} = Y^{a}(\Phi) + \overline{Y}^{a}(\bar{\Phi})
\end{align*}
where the $Y^{a}$ satisfy the \emph{complex} Killing equations
\begin{align*}
K_{r\bar{j}}\nabla_{i}Y^{r} + K_{i\bar{r}}\nabla_{\bar{j}}\overline{Y}^{\bar{r}} = 0
\end{align*}
This means that as we drag the geometry along the flowlines of the $L^{i}$, we consistently
preserve the complex-structure identifications Eq.~\eqref{complexstructure} in each of the
tangent spaces that we pass through. Consequently, every real Killing vector is also
holomorphic, and it does not matter whether we compute the vectors using the real or complex
Killing equations. In Appendix A we present the $L^{i}$ rewritten in terms of $S,T,Z$.

These vector fields $L^{i}$ now define seven symmetry directions of the K\"ahler manifold,
such that infinitesimal transformations along these directions leave the scalar-field
geometry (and hence interactions) invariant. Therefore, these vectors constitute a basis for
the Lie algebra of the isometry group, and their exponentiations will lead to the finite
symmetry transformations of the isometry group itself. The algebra can be determined by
investigating the commutation relations of the $L^{i}$, which are presented in Table 1
below.

\begin{table}[htbp] \label{table1}
\begin{center}
\begin{tabular}{|p{0.6cm}|c|c|c|c|c|c|c|}
\hline

&$L^{1}$  &$L^{2}$  &$L^{3}$  &$L^{4}$ &$L^{5}$ &$L^{6}$ &$L^{7}$ \\[6pt]
\hline
$L^{1}$ & 0 &$L^{2}$ &$-L^{3}$ &0 &$\frac{1}{2}L^{5}$ &0 &$-\frac{1}{2}L^{7}$\\[6pt]
\hline
$L^{2}$ & $-L^{2}$ &0 &$2L^{1}$ &0 &0 &0 &$-L^{5}$\\[6pt]
\hline
$L^{3}$ & $L^{3}$ &$-2L^{1}$ &0 &0 &$-L^{7}$ &0 &0\\[6pt]
\hline
$L^{4}$ &0 &0 &0 &0 &$-\frac{1}{2}L^{5}$ &$-L^{6}$ &$-\frac{1}{2}L^{7}$\\[6pt]
 \hline
$L^{5}$ &$-\frac{1}{2}L^{5}$ &0 &$L^{7}$ &$\frac{1}{2}L^{5}$ &0 &0 &$L^{6}$\\[6pt]
\hline
$L^{6}$ &0 &0 &0 &$L^{6}$ &0 &0 &0\\[6pt]
\hline
$L^{7}$ &$\frac{1}{2}L^{7}$ &$L^{5}$ &0 &$\frac{1}{2}L^{7}$ &$-L^{6}$ &0 &0\\[6pt]
\hline
\end{tabular}
\caption{\textit{Commutation relations of the Killing vector fields $L^{i}$}}
\end{center}
\end{table}

By inspection of the table we can see that $\{L^{1},L^{2},L^{3}\}$ form an
$sl(2,\mathbb{R})$ semisimple subalgebra, leaving over a solvable subalgebra $s \equiv
\{L^{4},L^{5},L^{6},L^{7}\}$. Hence, using the standard Levi-Malcev decomposition, the total
Lie algebra $g$ must read
\begin{align*}
g = sl(2,\mathbb{R}) \overrightarrow{\oplus} \: s
\end{align*}
where $\overrightarrow{\oplus}$ indicates a semidirect sum. To identify $s$ explicitly, we
first note that it possesses a three-dimensional, non-Abelian subalgebra
$h_{3}=\{L^{5},L^{6},L^{7}\}$ that can be uniquely identified as a Heisenberg algebra. To
see this, recall that the Heisenberg algebras $h_{p}$ are $p=2n+1$ dimensional
generalisations of the basic `uncertainty' relation, and take the form
\begin{align*}
[P_{i},P_{j}]=[Q_{i},Q_{j}]=[P_{i},C]=[Q_{i},C]=[C,C]=0 \quad,\quad
[P_{i},Q_{j}]=C\delta_{ij}
\end{align*}
They can be viewed as central extensions of the commutative algebra $\mathbb{R}^{2n}$ by
$\mathbb{R}$, with central term $C$. If we now relabel the generators as
$L^{5}=P,L^{6}=C,L^{7}=Q$ then the only non-trivial commutator for these three generators
becomes
\begin{align*}
[P,Q]=C
\end{align*}
as it should be for the Heisenberg algebra $h_{3}$. Using this information, we define
$D=-L^{4}$ and note that the total set of commutators for $s$ becomes
\begin{align*}
[P,Q]=C \quad, \quad [D,C]= C \quad,\quad [D,P]=\frac{1}{2}P \quad, \quad [D,Q]=\frac{1}{2}Q
\end{align*}
so that all the elements of the Heisenberg subalgebra $h_{3}$ are eigenvectors of the
additional generator $D$. The unique four-dimensional, solvable Lie algebra $s$ that
satisfies these requirements is the Lie algebra $su(1,2)/u(2)$ of the group $SU(1,2)/U(2)$,
i.e. the Borel (or Iwasawa) subalgebra of the Lie algebra $su(1,2)$~(see for example
Ref.~\cite{barberis97})~\footnote{Note that $SU(1,2)/U(2)$ is indeed a Lie group, since
$U(2)$ is a normal subgroup of $SU(1,2)$.}.

Therefore, we can now identify the total Lie algebra $g$ as
\begin{align}
g = sl(2,\mathbb{R}) \: \overrightarrow{\oplus} \: su(1,2)/u(2)
\end{align}
This happens to be isomorphic to one of the maximal parabolic subalgebras of
$sp(4,\mathbb{R})$, and has a unique lift via the exponential map to
\begin{align}~\label{group}
G = SL(2,\mathbb{R}) \ltimes SU(1,2)/U(2)
\end{align}
where $\ltimes$ defines the semidirect product (see for example Ref.~\cite{Kapovich}). This
group $G$ now defines the corresponding maximal parabolic subgroup of $Sp(4,\mathbb{R})$. To
find a suitable matrix description of this group, we note that a $4 \times 4$ matrix basis
for $g$ can obviously be found by considering a subset of the basis matrices for
$sp(4,\mathbb{R})$. Specifically, if $N^{i}$, $i=1 \ldots 10$, denotes the ten basis
matrices of $sp(4,\mathbb{R})$ (see Appendix B), then $N^{i}$, $i=1\ldots 7$, is a suitable
basis for $g$ with commutation relations identical to the Killing vector fields $L^{i}$.
Hence, by simple Lie exponentiation, one can show that the corresponding maximal parabolic
subgroup $G$ of $Sp(4,\mathbb{R})$ can be represented by real matrices of the form
\begin{align}~\label{groupmatrixrep}
\mathcal{G} =    \left(\begin{array}{cc}
                    A & B\\
                    C & D \\
                \end{array} \right)
\end{align}
with
\begin{align*}
A= \left( \begin{array}{cc}
             * &   0  \\
             * &   *
           \end{array} \right) \quad
B = \left( \begin{array}{cc}
             * &   * \\
             * &   *
    \end{array} \right) \quad
C = \left( \begin{array}{cc}
             * &   0 \\
             0 &   0
    \end{array} \right) \quad
D = \left( \begin{array}{cc}
             * &   *   \\
             0 &   *
    \end{array} \right)
\end{align*}
\begin{align*}
A^TD-C^TB=1_2 \quad, \quad A^TC \:, \: B^TD \quad \mathrm{symmetric}
\end{align*}
Here the $*$ indicates an entry that is not forced to be zero. This is a closed, symplectic
Lie group, and defines an arbitrary element of the isometry group $G$.

%=====================================SECTION 3=====================================================================

\section{The Group Action}

It is now useful to double-check this result, by showing that the structure of $G$ precisely
reproduces the finite group transformations represented by integrating the Killing vectors.
As a important corollary, we will also uncover the natural structure of the scalar-field
moduli space upon which the symmetry group $G$ acts.

To integrate the Killing vector fields, we first make the formal identification
\begin{align}\label{eq:killingflow}
L^{i} = \sum_{j} \frac{d\varsigma_{j}}{dc_{i}}\frac{\partial}{\partial\varsigma_{j}}
\end{align}
for each $i$, where $c_{i}$ is a set of parameters that measures how far one has moved along
the flowlines of the associated $L^{i}$, and $\varsigma =(\beta,\chi,\phi,\sigma,z,\nu)$ is
a row vector of moduli fields. By comparing coefficients on each side one finds that a set
of coupled differential equations must be satisfied, whose solution as a function of the
$c_{i}$ determines the finite, integrated transformations. These transformations are shown
below, with the equivalent complex results presented in Appendix C.

\begin{align}~\label{finitetrans}
L^{1}&: \quad \beta \rightarrow \beta + c_{1} && \chi \rightarrow \chi \:e^{c_{1}}
&&  z \rightarrow z \: e^{-c_{1}/2} && \nu  \rightarrow \nu \:e^{c_{1}/2}& \nonumber\\
L^{2}&: \quad e^{\beta} \rightarrow
\frac{e^{\beta}}{(1+c_{2}\chi)^{2}+\frac{1}{4}c_{2}^{2}e^{2\beta}}
 && \chi \rightarrow \frac{\chi(1+c_{2}\chi) + \frac{1}{4}c_{2}e^{2\beta}}{ %
\quad (1+c_{2}\chi)^{2}+\frac{1}{4}c_{2}^{2}e^{2\beta}}
 && \sigma \rightarrow \sigma- c_{2} \cdot 2q\nu^2 && z \rightarrow z + c_{2}\nu & \nonumber \\
L^{3}&: \quad \chi \rightarrow \chi + c_{3} &&  \sigma \rightarrow \sigma - c_{3} \cdot
2qz^{2} && \nu \rightarrow \nu + c_{3}z  &&   & \nonumber\\
L^{4}&: \quad \phi  \rightarrow \phi + c_{4} && \sigma \rightarrow \sigma \:e^{c_{4}} && z
\rightarrow z \: e^{c_{4}/2} && \nu \rightarrow \nu \:e^{c_{4}/2} & \nonumber \\
L^{5}&: \quad \sigma \rightarrow \sigma -4q\nu\cdot c_{5} &&  z \rightarrow z + c_{5} && &&
& \nonumber\\
L^{6}&: \quad \sigma \rightarrow \sigma +4qc_{6} &&   &&   &&   & \nonumber\\
L^{7}&: \quad \nu \rightarrow \nu + c_{7} &&   &&  && &
\end{align}

These transformations can be implemented as a matrix multiplication on the scalar fields,
where the latter have themselves been embedded as the entries of a symplectic $4 \times 4 $
matrix. To understand why this is possible, consider the following. It is well known that
the group $Sp(4,\mathbb{R})$ acts as a transitive group of diffeomorphisms on its own
solvable subgroup, which is the group manifold $\mathcal{M}=Sp(4,\mathbb{R})/U(2)$. These
diffeomorphisms are simply implemented by the multiplications
\begin{align}~\label{fullgroupaction}
\mathcal{V} \rightarrow \mathcal{U}\mathcal{V}\mathcal{S}
\end{align}
where $\mathcal{V} \in \mathcal{M}$, $S\in Sp(4,\mathbb{R})$ and $\mathcal{U} \in
Sp(4,\mathbb{R}) \cap O(4,\mathbb{R}) \cong U(2)$. That is, $Sp(4,\mathbb{R})$ acts by right
multiplications on $\mathcal{M}$, with a compensating local (field-dependent) left
multiplication by a member of the maximal compact subgroup $U(2)$. We can then equip
$\mathcal{M}$ with a metric invariant under $Sp(4,\mathbb{R})$, by taking a unique (up to
scaling) linear combination of the right-invariant one-forms on $\mathcal{M}$. This
transforms $\mathcal{M}$ into a Riemannian homogeneous space.

However, it is perfectly reasonable to consider a non-transitive set of diffeomorphisms
corresponding to right-multiplication on $\mathcal{M}$ \emph{only} by $G \subset
Sp(4,\mathbb{R})$, and define a metric on $\mathcal{M}$ that is \emph{only}
\mbox{$G$-invariant}. To see that this is consistent, consider that the unique Iwasawa
decomposition of $S \in Sp(4,\mathbb{R})$ takes the form $S=\mathcal{U}\mathcal{V}$, where
$\mathcal{V} \in Sp(4,\mathbb{R})/U(2)$, and $\mathcal{U} \in Sp(4,\mathbb{R}) \cap
O(4,\mathbb{R}) \cong U(2)$ is a member of the maximal compact subgroup. However, we also
have (up to an irrelevant discrete identification) the direct product decomposition
\begin{align*}
U(2) \cong U(1) \times SU(2)
\end{align*}
where the Abelian $U(1)$ factor is isomorphic to the maximal compact subgroup of $G$
(generated by the Killing vector $L^{3}$). By shuffling terms in the Iwasawa decomposition
one can then immediately see that \mbox{$\mathcal{M}=Sp(4,\mathbb{R})/U(2) \cong G/U(1)$},
so that the coset $G/U(1)$ is a group manifold isomorphic to $\mathcal{M}$. This means that
there is a well-defined representation of $G$ on $\mathcal{M}$ given by the action
\begin{align}~\label{ourgroupaction}
\mathcal{V} \rightarrow \mathcal{U}\mathcal{V}\mathcal{G}
\end{align}
where $\mathcal{V} \in Sp(4,\mathbb{R})/U(2)$, $\mathcal{G}\in G$ and $\mathcal{U} \in
U(1)$. This, in turn, is nothing but a consistent restriction of the transformations in
Eq.~\eqref{fullgroupaction} where we remove those diffeomorphisms that are not representable
as $G$ actions. Consequently, there must be some combination of the right-invariant
one-forms of $\mathcal{M}$ that defines a nonhomogeneous metric on $\mathcal{M}$ that is
only $G$ invariant.

Thus, it must be possible to represent the transformations Eq.~\eqref{finitetrans} in the
form Eq.~\eqref{ourgroupaction}, thus verifying that the isometry group is $G$ and the
underlying target space $\mathcal{M}$. To do this, consider gathering the scalar fields
$\beta,\chi,\phi,\sigma,z,\nu$ into the $4\times 4$ matrix $\mathcal{V}$, defined by

\begin{align}~\label{V}
\mathcal{V}&= %
e^{-\frac{1}{2}\beta N^{1}}e^{2\chi N^{2}}e^{\frac{1}{2}\phi N^{4}}e^{2\sqrt{q}\nu N^{5}}
e^{-(\sigma+4qz\nu)N^{6}}e^{\sqrt{q}zN^{7}} \nonumber \\ \nonumber \\
            &=\left(\begin{array}{cccc}
            e^{-\beta/2} & 0 & 2\chi e^{-\beta/2} & 2\sqrt{q}(\nu-\chi z)e^{-\beta/2} \\
            \sqrt{q}ze^{-\phi/2} & e^{-\phi/2} & 2\sqrt{q}\nu e^{-\phi/2} & (\sigma+2qz\nu)e^{-\phi/2}\\
            0 & 0 & e^{\beta/2} & -\sqrt{q}ze^{\beta/2}\\
            0 & 0 & 0 & e^{\phi/2}
            \end{array}\right)
\end{align}

This is nothing but an arbitrary element of $\mathcal{M}=Sp(4,\mathbb{R})/U(2)$, which in
turn is simply an arbitrary element of the solvable subgroup of $Sp(4,\mathbb{R})$ defined
by exponentiation of the solvable subalgebra generators $N^{i}$, $i=1,2,4,5,6,7$.
Consequently,  the six scalars are naturally related to the six group parameters of
$Sp(4,\mathbb{R})/U(2)$. We also take a matrix $\mathcal{G} \in G$ with constant entries as
follows
\begin{align*}
\mathcal{G} =\left(\begin{array}{cccc}
            e^{-c_1/2} & 0 & 2c_3 & 2\sqrt{q}c_7 \\
            \sqrt{q}c_5 & e^{-c_4/2} & 2\sqrt{q}(c_3c_5+c_7 \: e^{-c_4/2})\:e^{c_1/2} & 4qc_6\\
            \frac{1}{2}c_2 & 0 & (1+c_2c_3)\:e^{c_1/2} & \sqrt{q}(c_2c_7-c_5 \: e^{c_4/2})\:e^{c_1/2}\\
            0 & 0 & 0 & e^{c_4/2}
            \end{array}\right)
\end{align*}

Finally, if we define the function $u \equiv 2\left(e^{-c_1/2}+c_2\chi\right)e^{-\beta}$,
then the $U(1)$ ``compensator'' matrix $\mathcal{U}$ can be written~\footnote{To clarify:
$\mathcal{U}$ defines a subgroup of $Sp(4,\mathbb{R}) \cap O(4,\mathbb{R})$ that is
isomorphic to $U(1)$, and is represented by $4\times 4$ matrices.}
\begin{align*}
\mathcal{U} = \left(\begin{array}{cccc}
            \frac{u}{\sqrt{c_{2}^2+u^2}} & 0 & \frac{c_2}{\sqrt{c_{2}^2+u^2}} & 0 \\
            0 & 1 & 0 & 0\\
            -\frac{c_2}{\sqrt{c_{2}^2+u^2}}& 0 &  \frac{u}{\sqrt{c_{2}^2+u^2}} & 0\\
            0 & 0 & 0 & 1
            \end{array}\right)
\end{align*}
Overall, we then find that the total set of isometry transformations as defined in
Eq.~\eqref{finitetrans} can indeed be derived from the multiplications
\begin{align*}
\mathcal{V} \rightarrow \mathcal{U}\mathcal{V}\mathcal{G}
\end{align*}
Consequently, the isometry group is $G$ and the underlying manifold is
$\mathcal{M}=Sp(4,\mathbb{R})/U(2) \cong G/U(1) $. Note that the parabolic $G$ action is not
transitive, and so not every point in $\mathcal{M}$ can be connected to every other by a $G$
isometry (for a description of the fixed points see Ref.\cite{Goresky}). This stands in
stark contrast to the transitive isometry groups of homogeneous Riemannian spaces, and
suggests that there will be some obstruction to generating the general solution to the
equations of motion using the isometries alone.

If, instead, we wish to visualise these transformations in terms of the natural
\emph{complex} fields, then there is an equivalent and elegant alternative to the above
procedure. One can verify that there is a bijective map from $\mathcal{M}$ into the space of
complex, symmetric $2 \times 2$ matrices $\Psi$ with positive definite imaginary part. This
space is called the upper Siegel plane $SH_{2}=\{\Psi \in \mathrm{Sym}_{2}\mathbb{C} :
\mathrm{Im}(\Psi)>0\}$ (see for example \cite{Siegel, friedland1, friedland2}). Moreover,
there is a natural action of $Sp(4,\mathbb{R})$ (and so by restriction $G$) as a set of
fractional linear transformations over $SH_{2}$. Concretely, we first define the matrix
\begin{align*}
\Psi=   2i \left(\begin{array}{cc}
            S & \sqrt{q}Z  \\
            \sqrt{q}Z & T
            \end{array} \right)
\end{align*}
We then consider the transformation action
\begin{align*}
\left(\begin{array}{cc}
       \mathcal{A} & \mathcal{B}\\
       \mathcal{C} & \mathcal{D}\\
        \end{array} \right)
: \Psi \rightarrow
\left(\mathcal{A}\Psi+\mathcal{B}\right)\left(\mathcal{C}\Psi+\mathcal{D}\right)^{-1}
\end{align*}
where $\mathcal{A},\mathcal{B},\mathcal{C},\mathcal{D}$ are $2 \times 2$ real matrices
satisfying the symplectic conditions
\begin{align*}
\mathcal{A}^T\mathcal{D}-\mathcal{C}^T\mathcal{B}=1_2 \quad, \quad %
\mathcal{A}^T\mathcal{C},\; \mathcal{B}^T\mathcal{D} \quad \mathrm{symmetric}
\end{align*}
If we now restrict the action to those matrices of the form
\begin{align}~\label{g2rep}
\mathcal{G}_{2} =    \left(\begin{array}{cc}
                     \mathcal{A} & \mathcal{B}\\
                     \mathcal{C} & \mathcal{D} \\
                  \end{array} \right)
\end{align}
with
\begin{align*}
\mathcal{A} =  \left( \begin{array}{cc}
                    * &   *  \\
                    0 &   *
           \end{array} \right) \quad
\mathcal{B}  = \left( \begin{array}{cc}
                    * &   * \\
                    * &   *
    \end{array} \right) \quad
\mathcal{C}  = \left( \begin{array}{cc}
                    0 &   0 \\
                    0 &   *
    \end{array} \right) \quad
\mathcal{D}  = \left( \begin{array}{cc}
                    * &   0   \\
                    * &   *
    \end{array} \right)
\end{align*}
\begin{align*}
\mathcal{A}^T\mathcal{D}-\mathcal{C}^T\mathcal{B}=1_2 \quad, \quad
\mathcal{A}^T\mathcal{C}\:,\: \mathcal{B}^T\mathcal{D} \quad \mathrm{symmetric}
\end{align*}
then these produce the correct isometry transformations at the level of the complex fields.
One can verify explicitly that the matrices $\mathcal{G}_2$ do indeed define the maximal
parabolic subgroup $G$ again, but with a slightly different matrix representation to the one
specified by $\mathcal{G}$ in Eq.~\eqref{groupmatrixrep} (defined by a different embedding
in $Sp(4,\mathbb{R})$).

%================================SECTION 4===========================================================================

\section{Building the metric}

To come full circle, it should now be possible to build the $G$-invariant metric from the
invariant one-forms of the underlying manifold $\mathcal{M}$. To see this, recall that
$\mathcal{M}$ is a coset of the form $Sp(4,\mathbb{R})/U(2)$ with $U(2)$ a normal subgroup
of $Sp(4,\mathbb{R})$, and so $\mathcal{M}$ is a Lie group as well as a manifold.
Consequently, there is a set of canonically discriminated left-invariant vector fields on
$\mathcal{M}$, and an associated ``dual'' set of right-invariant one-forms. These dual
one-forms can be readily constructed from the matrix $\mathcal{V}$ in Eq.~\eqref{V}, by
forming the combination
\begin{align}
d\mathcal{V}\mathcal{V}^{-1} = \left(
\begin{array}{cccc}
-\frac{1}{2}d\beta & 0 & 2e^{-\beta}d\chi &
2\sqrt{q}e^{\frac{1}{2}\left(-\beta-\phi\right)}\left(d\nu-\chi dz\right)\\
\sqrt{q}e^{\frac{1}{2}\left(\beta-\phi\right)}dz & -\frac{1}{2}d\phi &
2\sqrt{q}e^{\frac{1}{2}\left(-\beta-\phi\right)}\left(d\nu-\chi dz\right) &
e^{-\phi}\left(d\sigma+4qz d\nu\right) \\
0 & 0 & \frac{1}{2}d\beta & -\sqrt{q}e^{\frac{1}{2}\left(\beta-\phi\right)}dz
\end{array}\right)
\end{align}
Each entry now defines a right-invariant one-form of the group $\mathcal{M}$, and these
forms are extremely useful for the following reason. The Killing vectors are left-invariant
vector fields, and one can verify that these fields correspond to transformations enacted by
infinitesimal \emph{right} multiplications on the Lie algebra of the group. Hence, if we
wish to assemble an invariant metric, this metric should be right-invariant and built from
the entries above. So let us define the following right-invariant one-forms
\begin{align*}
\theta_{1} &= \frac{1}{2}d\beta &  \theta_{2} &= e^{-\beta}d\chi\\
\theta_{3} &= \frac{1}{2}d\phi & \theta_{4} &= \frac{1}{2}e^{-\phi}
\left(d\sigma+4qzd\nu\right)\\
\theta_{5} &= \sqrt{\frac{q}{2}}e^{\frac{1}{2}\left(\beta-\phi\right)}dz & \theta_{6} &=
\sqrt{2q}e^{\frac{1}{2}\left(-\beta-\phi\right)}\left(d\nu-\chi dz\right)
\end{align*}
We now note that a choice of right-invariant metric corresponds to a choice of how to
combine these forms together. This is, implicitly, a choice of inner product on the Killing
vectors in the Lie algebra. For example, one simple choice is to take
\begin{align*}
ds^{2} = \sum_{i=1}^{7}\left(\theta_{i}\right)^2
\end{align*}
This defines the natural homogeneous, symmetric metric on $Sp(4,\mathbb{R})/U(2)$ and can be
derived from the K\"ahler potential
\begin{align}
K = -\ln\left[S+\overline{S}-q\frac{(Z+\overline{Z})^{2}}{T+\overline{T}}\right]
-\ln\left(T+\overline{T}\right) = - \ln \left[\mathrm{det}\:
\mathrm{Im}\left(\Psi\right)\right]
\end{align}
However, another equally valid choice -- which has an isometry group that is \emph{smaller}
than $Sp(4,\mathbb{R})$ -- is given by
\begin{align*}
ds^{2} = k\sum_{i=1}^{2}\left(\theta_{i}\right)^2 + \sum_{i=3}^{7}\left(\theta_{i}\right)^2
\end{align*}
where $k \neq 1$ and is a positive real constant. This derives from the K\"ahler potential
\begin{align}
K = -\ln\left[S+\overline{S}-q\frac{(Z+\overline{Z})^{2}}{T+\overline{T}}\right]
-k\ln\left(T+\overline{T}\right)
\end{align}
and cannot be written in a neat way in terms of the Siegel plane coordinate $\Psi$ because
the metric is nonhomogeneous. However, it is still a perfectly valid way of measuring
distances on the manifold, albeit with a reduced isometry group $G$. If we fix $k=3$ then
this is obviously the K\"ahler potential implied by the action Eq.~\eqref{4daction}, which
in turn is fixed by the structure of the higher-dimensional supergravity from which it
descends.

%======================================SECTION 5=====================================================================
\newpage%%%%%%%%%%%%%%%%%%%%%%%%%%%%%%%%%%%%%%%%%%%%%%%%%%%%%%%%%%%%%%%%%%%%%%%%%%%%%%%%%%%%%%%%%%%%%%%%%%%%%%%%%NEWPAGE
\section{Alternative derivation of the metric}

There is an alternative way to derive the metric, which allows one to rewrite it in a more
compact and elegant form. We noted previously that the inhomogeneous K\"ahler potential
Eq.~\eqref{kahlerpot} cannot be rewritten in terms of the natural complex coordinate $\Psi$
on $SH_{2}$. This is fundamentally because the obstructing factor of $3$ (which makes the
metric inhomogeneous) does not allow the terms to be gathered as the determinant of
$\mathrm{Im}\left(\Psi\right)$. However, one can bypass this obstruction by jumping up to
the higher-dimensional Siegel plane $SH_{4}\cong Sp(8,\mathbb{R})/U(4)$, using a
\emph{homogeneous} metric, and then killing off some fields. Consider, for example, defining
the complex coordinate
\begin{align*}
\Phi=   2i \left(\begin{array}{cc}
            \mathbf{S} &  \mathbf{Z}\\
            \mathbf{Z}^T & \mathbf{T}
             \end{array} \right)
\end{align*}
where $\mathbf{S},\mathbf{T},\mathbf{Z}$ are complex $2 \times 2$ matrices, with
$\mathbf{S},\mathbf{T}$ symmetric. If $\mathrm{Im}\left(\Phi\right)>0$ then $\Phi$ belongs
to the upper Siegel plane $SH_{4}$, defined as usual by $SH_{4}=\{\Phi \in
\mathrm{Sym}_{4}\mathbb{C} : \mathrm{Im}(\Phi)>0\}$. Moreover, there is a natural
homogeneous metric on $SH_{4}$ with a transitive group of $Sp(8,\mathbb{R})$ isometries,
which derives from the K\"ahler potential
\begin{align}~\label{sp8metric}
K =  - \ln \left[\mathrm{det}\: \mathrm{Im}\left(\Phi\right)\right]
\end{align}
Now consider restricting the 10 independent complex entries in $\Phi$, by demanding that it
reduce to the form
\begin{align*}
\tilde{\Phi} =   2i \left(\begin{array}{cccc}
                S         &  \sqrt{q}Z & 0   & 0 \\
                \sqrt{q}Z &  T         & 0   & 0  \\
                0         &  0         & T   & 0 \\
                0         &  0         & 0    & T \end{array} \right)
\end{align*}
This now defines a K\"ahler submanifold of $SH_{4}$, with an induced K\"ahler potential
given by the restriction of Eq.~\eqref{sp8metric} to this submanifold. This induced
potential takes the simple form
\begin{align}
K = - \ln [\:\mathrm{det}\: \mathrm{Im}(\tilde{\Phi})\:]
\end{align}
This is precisely the K\"ahler potential (and associated metric) that we have been analysing
in Eq.~\eqref{4daction}, now written in a more compact form. One can now readily determine
the isometry group of the metric, by looking for the subgroup $G \subset Sp(8,\mathbb{R})$
that preserves the form of $\tilde{\Phi}$ under fractional linear transformations. One can
verify that $G$ is represented by the following $Sp(8,\mathbb{R})$ matrices:
\begin{align*}
\mathcal{G}_3 = \left( \begin{array}{cc}
                    \mathfrak{A} & \mathfrak{B}  \\
                    \mathfrak{C} & \mathfrak{D}
                        \end{array} \right)
\end{align*}
where
\begin{align*}
\mathfrak{A} = \left( \begin{array}{clcc}
                            * &   *  & 0  & 0 \\
                            0 &   *  & 0  & 0  \\
                            0 &   0  & *  & 0  \\
                            0 &   0  & 0  & *
                        \end{array} \right) \quad
\mathfrak{B}= \left( \begin{array}{clcc}
                                * &   * & 0  & 0 \\
                                * &   * & 0  & 0  \\
                                0 &   0 & *  & 0 \\
                                0 &   0 & 0  & *
                            \end{array} \right)\quad
\mathfrak{C} = \left( \begin{array}{clcc}
                            0 &   0 &  0 & 0 \\
                            0 &   * &  0 & 0 \\
                            0 &   0 &  * & 0 \\
                            0 &   0 &  0 & *
    \end{array} \right) \quad
\mathfrak{D} = \left( \begin{array}{clcc}
                                * &   0  &  0 & 0 \\
                                * &   *  &  0 & 0  \\
                                0 &   0  &  * & 0  \\
                                0 &   0  &  0 & *
    \end{array} \right)
\end{align*}
with the additional constraints
\begin{align*}
\mathfrak{A}^T\mathfrak{D}-\mathfrak{C}^T\mathfrak{B}=1_4 \quad, \quad
\mathfrak{A}^T\mathfrak{C}\:,\: \mathfrak{B}^T\mathfrak{D} \quad \mathrm{symmetric}
\end{align*}

Due to the diagonal nature of the $2\times 2$ submatrices in the lower right blocks of
$\mathfrak{A},\mathfrak{B},\mathfrak{C},\mathfrak{D}$ above, the non-trivial group structure
is all captured by the $2\times 2$ submatrices in the upper left blocks. However, this set
of $2\times 2$ matrices defines the matrix representation $\mathcal{G}_{2}$ as given in
Eq.~\eqref{g2rep}. Consequently, $\mathcal{G}_2$ and $\mathcal{G}_{3}$ must be two different
representations of the same abstract group, which is none other than the maximal parabolic
group $G$. Thus, we can derive the nonhomogeneous metric from an ordinary homogeneous one,
by making appropriate field identifications and truncations. That is, the nonhomogeneous
character arises by imposing restrictions on the fields in an otherwise homogeneous
Riemannian space.

%===========================================SECTION 6===============================================================

\section{Conclusion}

We have investigated the effective $D=4$ description of heterotic M-theory, paying
particular attention to the isometries of the K\"ahler metric defined by the cosmological
fields. In doing so, we have presented our results in a variety of ways, and have had
occasion to switch between the real scalar fields and the complex superfields that they
group into. We now state the main results of this paper in the most convenient and
transparent form, in terms of the superfields $S,T,Z$. We discovered that the underlying
complex manifold is a submanifold of the upper Siegel plane $SH_{4}$, with each point of
this submanifold labelled by the matrix
\begin{align*}
\tilde{\Phi} =   2i \left(\begin{array}{cccc}
                S         &  \sqrt{q}Z & 0   & 0 \\
                \sqrt{q}Z &  T         & 0   & 0  \\
                0         &  0         & T   & 0 \\
                0         &  0         & 0    & T \end{array} \right)
\end{align*}
The K\"ahler potential can then be written in terms of this matrix as
\begin{align*}
K = - \ln [\: \mathrm{det}\: \mathrm{Im}(\tilde{\Phi}) \:]
\end{align*}
The isometries of the associated K\"ahler metric can be readily determined, by asking for
those transformations that preserve the form of $\tilde{\Phi}$ and act only as K\"ahler
transformations on the K\"ahler potential. The isometry group was found to be a group $G
\subset Sp(8,\mathbb{R})$, where $G$ is in fact a maximal parabolic subgroup of
$Sp(4,\mathbb{R})$ containing a set of $SL(2,\mathbb{R})$ $T$-duality transformations.
Specifically, the matrix $\tilde{\Phi}$ transforms as
\begin{align*}
\tilde{\Phi} \rightarrow
(\mathfrak{A}\tilde{\Phi}+\mathfrak{B})(\mathfrak{C}\tilde{\Phi}+\mathfrak{D})^{-1}
\end{align*}
where the matrices $\mathfrak{A},\mathfrak{B},\mathfrak{C},\mathfrak{D}$ are given by
\begin{align*}
\mathfrak{A} = \left( \begin{array}{clcc}
             * &   *  & 0  & 0 \\
             0 &   *  & 0  & 0  \\
             0 &   0  & *  & 0  \\
             0 &   0  & 0  & *
    \end{array} \right) \quad
\mathfrak{B} = \left( \begin{array}{clcc}
             * &   * & 0  & 0 \\
             * &   * & 0  & 0  \\
             0 &   0 & *  & 0 \\
             0 &   0 & 0  & *
    \end{array} \right)\quad
\mathfrak{C} = \left( \begin{array}{clcc}
             0 &   0 &  0 & 0 \\
             0 &   * &  0 & 0 \\
             0 &   0 &  * & 0 \\
             0 &   0 &  0 & *
    \end{array} \right) \quad
\mathfrak{D} = \left( \begin{array}{clcc}
             * &   0  &  0 & 0 \\
             * &   *  &  0 & 0  \\
             0 &   0  &  * & 0  \\
             0 &   0  &  0 & *
    \end{array} \right) \\
\end{align*}
subject to the symplectic conditions
\begin{align*}
\mathfrak{A}^T\mathfrak{D}-\mathfrak{C}^T\mathfrak{B}=1_4 \quad, \quad
\mathfrak{A}^T\mathfrak{C}\:,\: \mathfrak{B}^T\mathfrak{D} \quad \mathrm{symmetric}
\end{align*}

One can now use these $G$ isometries to make significant progress toward the general
solution to the equations of motion. In particular, one can build complicated new solutions
by applying these $G$ isometries to simpler, specific solutions (say where certain fields
are set to zero). In a companion paper we will derive several new classes of previously
unknown cosmological solutions to Eq.~\eqref{4daction}, including a solution where the M5
brane can reverse direction despite the absence of any explicit potentials.

It would also be interesting to clarify the origin of the exact $SL(2,\mathbb{R})$ symmetry
that we have found. For example, we know that in the absence of bulk M5 branes the K\"ahler
potential contribution for the $T$ modulus is just $-3\ln\left(T+\overline{T}\right)$, and
that this leads to an exact set of $T$-duality transformations. However, it is a non-trivial
feature that a modified version of this $T$-duality persists in the presence of an M5 brane.

%=============================================REFERENCES===========================================================
\newpage%%%%%%%%%%%%%%%%%%%%%%%%%%%%%%%%%%%%%%%%%%%%%%%%%%%%%%%%%%%%%%%%%%%%%%%%%%%%%%%%%%%%%%%%%%%%%%%%%%%%%%%%%%%NEWPAGE

\bibliographystyle{plain}

\section*{Appendix A}

The holomorphic Killing vector fields are given by
\begin{align*}
L^{1} &= ( T\partial_{T} + \frac{1}{2}\:Z\partial_{Z}) +
( \overline{T}\partial_{\overline{T}} + \frac{1}{2}\:\overline{Z}\partial_{\overline{Z}} ) \\
L^{2} &= \frac{i}{2}( qZ^{2}\partial_{S} + T^{2}\partial_{T}+ ZT\partial_{Z} ) -
\frac{i}{2}(q\overline{Z}^{2}\partial_{\overline{S}} +
\overline{T}^{2}\partial_{\overline{T}}+
\overline{Z}\overline{T}\partial_{\overline{Z}}) \\
L^{3} &= 2i\left(\partial_{T}-\partial_{\overline{T}}\right) \\
L^{4} &= ( S\partial_{S} +\frac{1}{2}\:Z\partial_{Z}) + (
\overline{S}\partial_{\overline{S}}+\frac{1}{2}\:\overline{Z}\partial_{\overline{Z}} )
\\
L^{5} &= \left( 2qZ\partial_{S} + T\partial_{Z} \right) +
\left( 2q\overline{Z}\partial_{\overline{S}} + \overline{T}\partial_{\overline{Z}} \right) \\
L^{6} &= 4qi\left(\partial_{S} - \partial_{\overline{S}}\right)\\
L^{7} &= -2i\left(\partial_{Z} -\partial_{\overline{Z}}\right)
\end{align*}

\section*{Appendix B}

A $4 \times 4$ matrix basis for the ten-dimensional Lie algebra $sp(4,\mathbb{R})$ is given
by
\begin{align*}
N^{1}&= \left(\begin{array}{cccc}
       1 & 0 &0 &0\\
       0 & 0 &0 &0\\
       0 & 0 &-1 &0\\
       0 &0 &0 &0 \end{array} \right)&
N^{2}&= \left(\begin{array}{cccc}
       0 & 0 &1 &0\\
       0 & 0 &0 &0\\
       0 & 0 &0 &0\\
       0 &0 &0 &0 \end{array} \right)&
N^{3}&= \left(\begin{array}{cccc}
       0 & 0 &0 &0\\
       0 & 0 &0 &0\\
       1 & 0 &0 &0\\
       0 &0 &0 &0 \end{array} \right)&
N^{4}&= \left(\begin{array}{cccc}
       0 & 0 &0 &0\\
       0 & -1 &0 &0\\
       0 & 0 &0 &0\\
       0 &0 &0 &1 \end{array} \right)\\
N^{5}&= \left(\begin{array}{cccc}
       0 & 0 &0 &1\\
       0 & 0 &1 &0\\
       0 & 0 &0 &0\\
       0 &0 &0 &0 \end{array} \right)&
N^{6}&= \left(\begin{array}{cccc}
       0 & 0 &0 &0\\
       0 & 0 &0 &-1\\
       0 & 0 &0 &0\\\
       0 &0 &0 &0 \end{array} \right)&
N^{7}&= \left(\begin{array}{cccc}
       0 & 0 &0 &0\\
       1 & 0 &0 &0\\
       0 & 0 &0 &-1\\
       0 &0 &0 &0 \end{array} \right)&
N^{8}&= \left(\begin{array}{cccc}
       0 & 0 &0 &0\\
       0 & 0 &0 &0\\
       0 & 0 &0 &0\\
       0 &1 &0 &0 \end{array} \right)\\
N^{9}&= \left(\begin{array}{cccc}
       0 & 1 &0 &0\\
       0 & 0 &0 &0\\
       0 & 0 &0 &0\\
       0 &0 &-1 &0 \end{array} \right)&
N^{10}&= \left(\begin{array}{cccc}
       0 & 0 &0 &0\\
       0 & 0 &0 &0\\
       0 & 1 &0 &0\\
       1 &0 &0 &0 \end{array} \right)
\end{align*}

\section*{Appendix C}

The finite superfield transformations corresponding to $L^{1} \ldots L^{7}$ are
\begin{align*}
L^{1}&: \quad S = S_{0} &&  T = T_{0}\:e^{c_{1}} &&  Z = Z_{0}\:e^{c_{1}/2}&\\
L^{2}&: \quad S = S_{0} + \frac{qZ_{0}^{2}}{T_{0}}\left(\frac{ic_{2}T_{0}}{2-ic_{2}T_{0}}\right) %
&&  T = \frac{2T_{0}}{2-ic_{2}T_{0}} %
&&  Z = \frac{2Z_{0}}{2-ic_{2}T_{0}}&\\
L^{3}&: \quad S = S_{0} && T = T_{0} +2ic_{3} && Z = Z_{0} &\\
L^{4}&: \quad S = S_{0}\:e^{c_{4}} && T = T_{0} && Z = Z_{0}\:e^{c_{4}/2}&\\
L^{5}&: \quad S = S_{0}  + 2qZ_{0}c_{5}+ qT_{0}c_{5}^{2} && T = T_{0} &&
Z = Z_{0}+c_{5}T_{0}&\\
L^{6}&: \quad S = S_{0}+4qic_{6} && T = T_{0} && Z = Z_{0}&\\
L^{7}&: \quad S = S_{0} && T = T_{0} && Z = Z_{0}-2ic_{7}&\\
\end{align*}

\end{document}